# Hybridization of colloidal handlebodies with singular defects and topological solitons in chiral liquid crystals

Jun-Yong Lee[a], Asha Kumari[a], Ye Yuan[a], Mykola Tasinkevych[ab], and Ivan I. Smalyukh[*acde]

[a]International Institute for Sustainability with Knotted Chiral Meta Matter, Hiroshima University, 1-3-1 Kagamiyama, Higashi-Hiroshima, Hiroshima 739-8526, Japan. Email: ivan.smalyukh@colorado.edu

[b]Department of Physics and Mathematics, School of Science and Technology, Nottingham Trent University, Clifton Lane, Nottingham NG11 8NS, United Kingdom

[c]Department of Physics, University of Colorado, Boulder, CO 80309, USA.

[d]Department of Electrical, Computer, and Energy Engineering, Materials Science and Engineering Program and Soft Materials Research Center, University of Colorado, Boulder, CO 80309, USA.

[e]Renewable and Sustainable Energy Institute, National Renewable Energy Laboratory and University of Colorado, Boulder, CO 80309, USA

* Correspondence to: ivan.smalyukh@colorado.edu

Abstract

Topology can manifest itself in colloids when quantified by invariants like Euler characteristics of nonzero-genus colloidal surfaces, albeit spherical colloidal particles are most often studied, and colloidal particles with complex topology are rarely considered. On the other hand, singular defects and topological solitons often define the physical behavior of the molecular alignment director fields in liquid crystals. Interestingly, nematic liquid crystalline dispersions of colloidal particles allow for probing the interplay between topologies of surfaces and fields, but only a limited number of such cases have been explored so far. Here, we study the hybridization of topological solitons, singular defects, and topologically nontrivial colloidal particles with the genus of surfaces different from zero in a chiral nematic liquid crystal phase. Hybridization occurs when distortions separately induced by colloidal particles and LC solitons overlap, leading to energy minimization-driven redistribution of director field deformations and defects. As a result, hybrid director configurations emerge, combining topological features from both components. We uncover a host of director field configurations complying with topological theorems, which can be controlled by applying electric fields. Rotational and translational dynamics arise due to the nonreciprocal evolution of the director fields in response to alternating electric fields of different frequencies. These findings help define a platform for controlling topologically hyper-complex colloidal structures and dynamics with electrically reconfigurable singular defects and topological solitons induced by colloidal handlebodies.

**1. Introduction**

Colloids typically consist of nano-to-micrometer-sized particles suspended in most commonly an continuous medium, such as water, and are of great significance in both fundamental science and practical applications [1-4]. The physical properties and behavior of colloidal systems can be tuned by external stimuli, including temperature change, electromagnetic field, etc., as well as intrinsic characteristics such as the size distribution and geometric shape of constituting particles. While spherical colloidal particles have been extensively studied due to their simple geometry, abundance, and ease of both preparation and analysis [5-13], research on topologically complex particles [14-20], such as those with nonzero-genus [14-16], remains limited. These complex colloids introduce possibilities for examining molecular-colloidal and surface-field interactions mediated by topological properties, providing insights into how topological features can govern system behaviors.

Unlike classical colloidal systems in isotropic media, colloidal behaviors may be further enriched by anisotropic suspension media such as liquid crystals (LCs), which also exhibit topological features in their collective molecular orientation field [5-20] The director, n(r), which represents the spatial distribution of local average molecular orientation of LC molecules [21, 22] may appear uniform and constant n(r)=n0 when undisturbed and defined by surface boundary conditions or deformed when perturbed by colloidal inclusions or external stimuli. Among the various types of director deformations, topologically nontrivial ones, such as singular defects [14-16, 23-27], where the director is undefined, and topological solitons [28-38], which feature spatially localized continuous but topologically nontrivial field deformations with particle-like properties, are of particular interest. Solitonic topological structures recently discovered in colloids [39], LCs [31-33] and polymerized systems [40] include torons, hopfions, skyrmions, helikontons, various cholesteric fingers and their loops or periodic arrays, etc., which can be classified by homotopy theory, similar to how singular defects are classified. In addition to being of interest to fundamental science, these topological structures hold potential for applications ranging from information displays to electro-optic and photonic devices.

Colloidal particles suspended in LCs induce director deformations determined by the surface characteristics like genus and boundary conditions for n(r) [8, 9, 14-16, 41-44]. For topological defects, studies have found [14-16] that particle surface topologies dictate the total topological invariants, although their exact structures, location, and surrounding director configuration are determined by the minimization of the system's total free energy. However, despite extensive investigations into solitonic structures in LCs [28-38], including their interactions, collective schooling behaviors, and interplay with introduced colloids, research exploration of the interplay between colloidal particles and solitons of nontrivial topology in LCs remains limited.

In this work, we study the hybridization of colloidal particles with singular defects and topological solitons in confinement-frustrated and unwound chiral nematic liquid crystals (CNLCs), where topological theorems relate invariants characterizing topology of two-dimensional colloidal surfaces and that of singular defects, co-hosted along with various solitonic structures. We find that the chiral solitonic features can develop beneath/above or around the handlebody particles. Furthermore, topological solitons such as torons interact with colloidal handlebodies via elasticity-mediated colloidal interactions minimizing the energetic costs of elastic deformations in the director field around both particles and torons, leading to self-assembly into bound states. Electric fields allow for reconfiguring these hybridized states, e.g., by shrinking or expanding the solitonic features around particles. Periodically oscillating electric fields applied to such samples modulate the coronae of elastic deformations in a nonreciprocal way, leading to squirming-like and other types of translational and rotational motions of the colloidal particles. Although the behavior of this system is complex, we demonstrate how such dynamics can be controlled by external stimuli like electric fields. We then discuss how the interplay of topologies of colloidal surfaces and both singular and solitonic director fields can define a platform for new means of controlling colloidal self-assembly.

## 2 Methods

### 2.1 Experimental

Two types of LC mixtures were prepared. The first mixture consisted of 4-Cyano-4'-pentylbiphenyl (5CB) and the chiral dopant (*S*)-4-Cyano-4'-(2-methylbutyl) biphenyl (CB15) to create a chiral nematic phase, resulting in cholesteric pitch *p*=5-20 μm with 2.8 wt% - 0.7 wt% of CB15. For the second mixture, 5CB and 4-(heptane-1,7-diyl)-dibiphenyl-4-carbonitrile (CB7CB) were combined at 125°C, at which the mixture exhibits the isotropic phase, under active stirring to ensure homogeneity. This mixture, containing 70 wt% of 5CB, exhibited the nematic phase at room temperature. A small amount (1.5 wt% - 0.4 wt%) of CB15 was added to this nematic mixture to produce the right-handed CNLC phase. Both mixtures exhibit positive dielectric anisotropy in the studied frequency range. All experiments, except those shown in Fig. 7(a–c) and Fig. 7(f–h), were conducted using the first mixture. For these specific experiments, the second mixture, containing additional CB7CB, was used to modify the bend elastic constant of the medium and stabilize the solitonic structures [45,46]. Additionally, 0.01 wt% of the fluorescent dye N, N-Bis(2,5-di-tert-butylphenyl)-3,4,9,10-perylenedicarboximide (BTBP) was added to both mixtures for confocal fluorescence microscopy imaging. Due to the anisotropic nature of the used fluorescent dye molecules, which follow the LC director, regions where the director aligns parallel to the linear polarization appear brighter under the microscope, facilitating the analysis of the director alignment.

Topological handlebody particles, with nonzero-genus and a thickness of approximately 500 nm (Supplementary Fig. S1), were fabricated on a silicon wafer using photolithographic techniques detailed in previous reports [15,16]. The particles were then released from the substrate using a 2 wt% sodium hydroxide aqueous solution by dissolving the aluminum sacrificial layer under ultrasonication. The particles were allowed to sediment within a centrifugation vial overnight, and the supernatant was replaced by deionized (DI) water. This process was repeated three times to remove excess sodium hydroxide. Without further surface functionalization, the particles exhibited planar surface anchoring when suspended in the used LCs [16]. To achieve homeotropic anchoring boundary conditions, the particles were first immersed in a 1 wt% aqueous solution of dimethyloctadecyl[3-(trimethoxysilyl)propyl]ammonium chloride (DMOAP) for 30 min and then washed in ethanol to remove excess DMOAP following the same procedure to remove sodium hydroxide as described before. After the cleaning and functionalization, the water or ethanol-based suspension containing the handlebodies was combined with the LC mixtures, which were then kept at 70°C for a day to remove the solvents and obtain particle-LC chiral nematic colloidal dispersion.

Samples for optical imaging were prepared by infiltrating the particle-LC dispersion into cells made of indium-tin-oxide (ITO)-coated glass plates, pre-cleaned by sequential ultrasonication in DI water, acetone, and ethanol for 15 min each. Both cell substrates were coated with DMOAP or lecithin (both from TCI) to achieve homeotropic alignment of the director. For DMOAP treatment, the substrates were immersed in an aqueous solution containing 1 wt% DMOAP for 30 min, subsequently rinsed with DI water to remove excess DMOAP, and dried at 100 °C for 30 min. For lecithin treatment, a 1 wt% lecithin ethanol solution was dropped onto the substrate and wiped off using a low-lint tissue. The treated substrates were then assembled into cells with a gap of *d*=5–20 μm defined by glass spacers to match with the cholesteric pitch such that *d/p*~1 or smaller. Electrodes were soldered on the ITO plates and connected to a function generator (AFG-3031, GW Instek) with a square wave output to apply an electric field on the sample, allowing control over the LC far-field director, soliton formation, and dynamic behavior of the handlebody particles at various frequencies.

Microscopic imaging was performed using a multifunctional microscope (FV3000, Olympus) equipped with bright-field, polarized optical, and fluorescence confocal polarizing microscopy (FCPM) capabilities to image and analyze LC director distributions and colloidal particle behavior. Particle spatial trajectories and orientation angles were tracked using video microscopy. High-resolution videos with 10-20 frames per second were analyzed with ImageJ (freeware for the National Institute of Health), utilizing particle tracking plugins to extract the position and the angle of individual particles in each frame. In the case of FCPM imaging, the strength of fluorescence emission depends on the angle between the laser excitation light's polarization and the orientation of the transition dipole moments of fluorescent dye molecules (BTBP). BTBP is an anisotropic fluorescent dye that aligns parallel to the local LC director. Due to this alignment, the fluorescence intensity depends on the polarization direction of the excitation light, reaching its maximum when the excitation polarization is parallel to the local director. Therefore, by introducing a small concentration of BTBP into the LC mixture and acquiring confocal images while varying the polarization direction of the excitation light, we can identify the director field distribution [47].

**2.2 Numerical simulations**

We model the cholesteric liquid crystal by using Landau-de Gennes free energy approach, based on a symmetric traceless tensor order parameter, $Q_{ij}(i,j = 1, \ldots, 3)$, which has five independent components. The Landau-de Gennes free energy $F_{LdG}$ of a chiral nematic may be written in the following form [21]

$$F_{LdG} = \int_V \left( \begin{array}{c} aQ_{ij}^2 - bQ_{ij}Q_{jk}Q_{ki} + c(Q_{ij}^2)^2 + \frac{L_1}{2}\partial_k Q_{ij} \partial_k Q_{ij} \\ + \frac{L_2}{2}\partial_j Q_{ij} \partial_k Q_{ik} + \frac{4\pi}{p}\epsilon_{ijk} Q_{il} \partial_j Q_{kl} \end{array} \right) dV, \quad (1)$$

where summation over repeated indices is assumed, and $\epsilon_{ijk}$ is the Levi-Civita symbol. In equation (1), the parameter $a$ is assumed to depend linearly on temperature $T$ with $a(T) = a_0(T - T^*)$, where $a_0$ is a material-dependent constant, and $T^*$ is the supercooling temperature of the isotropic phase. Phenomenological parameters *b* and *c* are assumed temperature independent, elastic parameters $L_1$ and $L_2$ can be expressed in terms of the Frank-Oseen elastic constants [48], and $p$ is the cholesteric pitch. We describe finite homeotropic anchoring, with a positive anchoring strength coefficient $W$, at the surface of a handlebody particle by using the following surface anchoring free energy

$$F_s = W \int_{\partial V} (Q_{ij} - Q_{ij}^s)^2 \, dS, \quad (2)$$

where $Q_{ij}^s = \frac{3Q_s}{2}(\nu_k \nu_j - \frac{\delta_{kj}}{3})$, $\delta_{ij}$ is the Kronecker delta symbol, $\boldsymbol{\nu}$ is the unit outward normal vector to the particle surface [49], and $Q_s$ is the preferred surface scalar order parameter, which we set for simplicity to the value of the scalar order parameter in the bulk nematic phase $Q_b = b/8c \, (a + \sqrt{1 - 64 \, ac/(3b^2)})$. The total free energy $F = F_{LdG} + F_s$ is then minimized numerically by employing the adaptive mesh finite elements method, as described in more detail in Ref. 48.

We assume a liquid crystal domain size $V = \{0 \leq x, y \leq L, 0 \leq z \leq h\}$, and use as the initial conditions for the numerical minimization of $F$ simple uniaxial Ansatz for the order parameter corresponding to a toron [50]. We apply periodic boundary conditions in the $x$ and $y$ directions, and adopt at the surfaces $z = 0$ and $z = h$ weak homeotropic anchoring given by equation (2) with the anchoring strength $W_s$.

We introduce the dimensionless temperature $\tau = 24a(T)c/b^2$ and the correlation length $\xi = 2\sqrt{2c(3L_1 + 2L_2)}/b$ at the nematic-isotropic transition. The cholesteric phase is stable for $\tau < \tau_{CI}$, where the cholesteric-isotropic coexistence temperature is given by Ref. 21

$$\tau_{CI} = \begin{cases} \frac{1}{2}\left(1 + (q_0\xi)^2 + \left(1 + \frac{1}{3}(q_0\xi)^2\right)^{\frac{3}{2}}\right), & q_0\xi \leq 3 \\ (q_0\xi)^2, & q_0\xi > 3, \end{cases} \quad (3)$$

where $q_0 \equiv \frac{2\pi}{p}$. In our calculations, we use $a_0 = 0.044 \times 10^6$ J K m$^{-3}$, $T^* = 307$ K, $a(T) = 0.02 \times 10^6$ J m$^{-3}$, $b = 0.816 \times 10^6$ J m$^{-3}$, $c = 0.45 \times 10^6$ J m$^{-3}$, $L_1 = 6 \times 10^{12}$ J m$^{-1}$, and $L_2 = 12 \times 10^{12}$ J m$^{-1}$, which are typical values for 5CB [51]. For these values of the model parameters, the bulk correlation length $\xi \approx 15$nm, which set the scale for the spatial extension of inhomogeneous regions and the cores of topological defects. We set the homeotropic anchoring strength $W = 10^{-3}$ J m$^{-2}$ for the surface of the handlebody particle and $W_s = 10^{-4}$ J m$^{-2}$ for the surfaces $z = 0$ and $z = h$. These values correspond to the surface anchoring strengths of interactions for surfaces coated with common surfactants [52-54]. We also set $p = 1$ μm, $h = 1$ μm, and $L = 8$ μm. The $g = 1$ handlebody particle geometry is as follows: the outer and inner radii are 0.6 μm and 0.3 μm, and the thickness of the handlebody tube is 0.3 μm.

We employ the Open Source Gnu Triangulated Surface (GTS) library [55] to generate triangulated surfaces representing the $g = 1$ handlebody particles. Next, the discretization of the liquid crystal domain $V$ was performed by using a quality tetrahedral mesh generator [56], which supports adaptive mesh refinement. Finally, the sum of the discretized functionals (1) and (2) is minimized numerically by using the M1QN3 optimization routine developed at INRIA [57], which implements a limited-memory quasi-Newton method of Nocedal [58].

## 3 Results

### 3.1 Hybridization of Solitonic Structures and Handlebody Particles

Colloidal inclusions in LCs induce singular defects within the director field, governed by the constraint defined by the surface topology of the particles. For instance, the total topological charge of bulk defects induced by colloids with perpendicular boundary conditions, $\Sigma m = \pm(1-g)$, and total winding number of surface defects, $\Sigma s = \pm(2-2g)$ on colloidal surfaces under planar boundary conditions, depend on $g$, the genus of the handlebody particle [15-16, 59]. Particle-like solitons in LCs with much complex director distortions also require additional topological defects to be embedded in the uniform background field, i.e., net zero of all topological charges of accompanying singular defects [30,59]. However, the exact amount and distribution of the defects are determined by the minimization of the free energy cost associated with the director distortions, which include continuous deformations and singular defects of the entire system. When these distortions separately induced by colloidal particles and LC solitons overlap, energy minimization again drives the re-distribution of director field deformations and defects, leading to hybrid director configurations combining topological features from both components while still following the topological constraints.

In this work, we report such hybridization in CNLC cells where alternating electric fields and the intrinsic chiral nature of the host medium facilitate the formation of solitons and the resulting hybrid structures with handlebody particles. When a handlebody particle is positioned appropriately during the formation of solitons, hybrid structures emerge with solitonic features above/beneath the particle (Figs. 1 and 3d-j) or surrounding it (Figs. 2 and 3a-c).

A simple example of such hybridization occurs between a toron and a $g=1$ handlebody particle (the torus-shaped particle) with planar anchoring suspended in a 5-μm-thick CNLC cell with homeotropic boundaries and a chiral pitch of 10 μm (Fig. 1). Bright-field (Fig. 1a) and polarized optical microscopy (POM) images (Fig. 1b) reveal a clearly distinct solitonic region encapsulating the particle. FCPM imaging (Fig. 1c) confirms its toron-like structure through its rotational symmetry (Supplementary Fig. S2b) and double-twist director field, as revealed by the vertical (Fig. S2a) and horizontal (Fig. S2b) cross-sectional images. FCPM imaging (Fig. S2) further reveals that the particle is asymmetrically positioned along the *z*-direction within the cell's vertical cross-section containing the toron.

The toron-torus soliton-colloidal hybrid structure can be understood by leveraging knowledge about the isolated and separate director configurations formed around a $g=1$ handlebody particle with planar anchoring and that of a toron. The resemblance between the director configurations along the midplane of a toron [31] and that observed around a standalone $g=1$ handlebody particle with planar anchoring [16] (Fig. 1d, e) suggests that the particle can embed into the interior of the toron. Within this configuration, no additional defects are induced by the handlebody pursuant to the topology constraint since $s=0$ when $g=1$, while the resulting hybridized toron retains two defects, a hedgehog defect, and a disclination loop, located along the central axis above and below the skyrmion-tube-like structure. This hypothesis is supported by FCPM images showing that the toron features a point defect at one end of the skyrmion tube fragment (Fig. S2b), while the opposite end, where the particle resides, contains a disclination loop (Fig. S2a, c), topologically equivalent to an elementary $\pi_2(\mathbb{S}^2/\mathbb{Z}_2)$ hedgehog defect and satisfy the hedgehog charge conservation rule (Fig. 1f). FCPM imaging, in which bright regions correspond to areas where the LC director aligns with the polarization direction of the excitation light, clarifies how these defects appear. In Fig. S2b, the small central region remains dark across four different polarization directions separated by a 45° interval, indicating the presence of a point defect. In Fig. S2c, although the disclination loop itself is not immediately visible, the reconstructed field topology [31] of torons, along with the open configuration observed in Fig. S2a, suggests that the point defect transitions into a loop. This interpretation, the transition of a point defect into a disclination loop, is further supported by comparing the microscope observations of the hybridized states with that of a standalone toron (Figs. 1g-j). Near the lower-substrate plane, a point defect is observed in both the hybridized and standalone torons (Fig. 1h). However, near the upper substrate, only the standalone toron maintains a point defect, while the hybridized structure shows an open disclination loop configuration around the particle (Fig. 1j). This suggests that hybridization can alter the solitonic structure while satisfying the topological invariant.

Different solitons feature varying forms of hybridization. Shown in Fig. 2, a finger-loop-like configuration featuring $2\pi$ director rotations in its midplane cross-section accommodates a $g=1$ handlebody particle with homeotropic surface anchoring in its interior. A POM image (Fig. 2a) reveals that a wide and distinct solitonic region encircles the particle, which contracts and spatially localizes as the applied voltage increases (Fig. 2b). The particle's positioning within the solitonic structure indicates a distinct hybrid configuration, different from that observed for the toron in Fig. 1, as also seen from the FCPM images (Figs. 2c-e). FCPM images intersecting the particle reveal an axially symmetric director configuration around the particle (Fig. 2c). At a slightly higher focal plane, a brightness pattern that exhibits two maxima as it extends radially from the center (Fig. 2e) is observed, consistent with the $2\pi$ director rotation along the radial directions. Together with the cross-sectional images of FCPM (Fig. 2d), these observations confirm that the surrounding solitonic structure adopts a finger-loop-like nonsingular configuration (Fig. 2g) [31] A schematic diagram of this hybrid configuration is presented in Figs. 2f and 2g, constructed based on the experimental observations. The abrupt director deformation between he $2\pi$ twist and the outer boundary of the handlebody (Fig. 2g) can be explained by a half-integer disclination loop with a local winding number of -1/2, which introduces a discontinuity. In this case, the winding number of -1/2 does not directly correspond to topological charge of -1. Due to the non-polar symmetry of the nematic director ($\mathbf{n}=-\mathbf{n}$), the sign of the hedgehog charge is not always intrinsically determined, but depends on the chosen vectorization. Reversing the vectorization inverts the sign of the topological charge but does not alter the relative charge between the defects, and the total charge must still satisfy the topological theorem. Therefore, the $g=1$ handlebody with perpendicular boundary condition induces defects of zero net topological charge following a topological theorem, $\Sigma m=\pm(1-g)$ [15], resulting in oppositely charged disclination loops close to the inner and outer facets of the particle (Fig. 4e). This configuration also ensures the overall $\pi_2(\mathbb{S}^2/\mathbb{Z}_2)$ hedgehog charge conservation [15].

Similar yet still different hybridization can be observed between solitonic structures and particles of more complex topology, e.g., a $g=3$ handlebody with planar or homeotropic anchoring (Fig. 3). With planar anchoring at the particle surface, the hybridized configuration features a solitonic loop which wraps around the $g=3$ handlebody (Figs. 3a, b, Supplementary Fig. S3a). A reconstructed director field based on FCPM images [31, 32] further reveals the configuration around the particle (Fig. 3c). In contrast, for a particle with homeotropic anchoring, the solitonic structure encloses the particle within itself. Bright-field (Fig. 3d) and POM images (Fig. 3e) show a distinct solitonic region surrounding the particle, accompanied by a hedgehog point defect in one of the particle's holes.

Electric fields can mediate hybridization between handlebody particles and solitons. The dynamic process of soliton formation under an electric field, particularly of the concentric finger-loop-like feature here, begins with elongated finger-like structures that contract over time into the stable localized solitons, as shown in Supplementary Fig. S4. Moreover, electric fields reconfigure the hybrid structure. For example, when an AC electric field of peak-to-peak voltage $U_{pp}$=1–15 V at 10 kHz is applied, the solitonic region surrounding the $g$=3 handlebody with homeotropic anchoring contracts as the field amplitude is increased, and the hedgehog defect moves from the hole's center toward the particle's center (Fig. 3g). Upon further increasing the voltage up to $U_{pp}$=15 V, the point defect opens up into a half-integer disclination loop, which remains stable even after the electric field is reduced back to $U_{pp}$=5 V (Figs. 3h, i). A similar transition, between a point defect and a disclination loop, arises during regeneration of hybridized configuration between a solitonic structure and a $g$=2 handlebody by turning the electric field off and on. (Supplementary Fig. S5). This configurational transition is driven by the minimization of the total free energy [15] favoring disclination loops at high electric fields while preserving the hedgehog topological charge.

**3.2 Elasticity-Mediated Interactions**

While the spontaneous hybridization between colloidal particles and solitons may lead to the emergence of stable composite structures, LC elasticity-mediated interactions between such initially separated entities can also occur (Figs. 4-6). In such cases, the handlebody particles and solitons interact through minimizing the energetic costs of the elastic distortions of the director field within the surrounding CNLC medium. For example, a toron generated by applying an electric field of $U_{pp}$=1.2 V at 1 kHz in a homeotropic cell with a thickness of 5 μm, and a $g$=1 handlebody particle are observed to approach each other over ~30 s when released at the initial separation of ~9 μm (Figs. 4a-b). The two objects eventually come into close contact, with the toron slightly deformed in shape, suggesting asymmetric influence from the director distortion induced by the handlebody particle.

Given their colloidal nature and the relatively high viscosity of the surrounding LC medium [61], the system's dynamics can be described under the low Reynolds number approximation, where inertia effects can be neglected. The viscous Stokes drag force, $\mathbf{F}_d$=-$b$(d$R$/d$t$), is thus balancing the attractive force originating from the perturbed LC director field, where $b$ and $R$ represent the drag coefficient and the center-to-center distance between the toron and the particle, respectively. Here, for simplicity, the drag coefficient $b$ is approximated by $6\pi\eta r$, i.e., the expression for an equivalently-sized spherical particle in a Newtonian fluid, where $\eta$=69 mPa s [61] and $r$=2 μm are, respectively, the viscous coefficient of 5CB and the characteristic size of both the toron and the handlebody particle. The approaching relative speed increases as the interacting objects come closer (Fig. 4c), indicating a stronger attractive elasticity-mediated force. Integration of the force over distance yields a binding energy of ~1200 $k_BT$, where $k_B$ is the Boltzmann constant and $T$=300 K the room temperature. This binding energy arises from the minimization of the LC elastic free energy upon bringing the toron and the handlebody particle to close contact. The elastic energy minimization leads to the overlap of the regions with strong director deformations, thereby reducing the system's free energy. It should be noted that this simple calculation may slightly overestimate the binding energy, as the shape of the handlebody particle is more like a disc rather than a sphere, suggesting a smaller drag coefficient than that assumed in the calculation. The observed attractive toron-handlebody interaction is consistent with the complementary lock-key-like director deformations induced by the toron and the handlebody with homeotropic surface anchoring at their circumferential space (Fig. 4d, e), which also propagate beyond the close vicinity of the two. Therefore, it is energetically more favorable for them to approach each other while eliminating extra deformations in-between, resulting in an effective attraction. Such elasticity-mediated interactions can be described within the framework of nematostatics, where deformations of the director field can be understood in analogy to that of electrostatic charge distributions [8].

Numerical calculations of the director configurations and associated free energy costs versus the distance between a toron structure and a $g$=1 handlebody generally support the above analysis and qualitative reasoning (Fig. 5). Numerically calculated elastic contribution from the Landau-de Gennes free energy as a function of the separation $R$ between a toron and a $g$=1 handlebody with homeotropic anchoring is generally consistent with experimental findings, revealing an attractive interaction for $R > 1.5p$, where $p$ is the cholesteric pitch. In this separation regime, the toron and handlebody are well separated, as shown in configurations marked (3) and (4) in Fig.5. In contrast, at smaller distance, the handlebody and the toron start overlapping (see configurations (1) and (2) in Fig. 5), forming hybrid toron-handlebody configurations. At $R$~1.4$p$ or smaller, the interaction between the toron and the handlebody becomes repulsive, revealing the regime of interactions not probed in experiments. Thus, overall, the numerical simulation findings reveal the existence of two branches in the free energy as a function of $R$. To obtain the repulsive free energy branch, one would need to use laser tweezers or other means to squeeze the particle-induced and toron's director deformations, which is outside the scope of our present study. Nevertheless, the numerical modelling reveals how a singular disclination loop, which localizes in the proximity of the colloidal handlebody's exterior, can hybridize with one of the small loops of the toron quasiparticle (corresponding to the internal structure of the point defects), thus forming a different hybridized state. This bound state of the defect-hybridized with disclination loop's inter-connection is therefore yielding an energetic binding behavior different from that seen for torons and tori with disconnected defect loops, where binding is associated purely with **n(r)**-deformations.

An even more complex form of elasticity-mediated interaction is observed between a toron and a hybrid structure consisting of a $g=3$ handlebody particle decorated by solitonic structures (Fig. 6). In this case, the handlebody particle with planar anchoring is oriented with its plane perpendicular to the confining surfaces (Fig. 6a). A standalone toron is attracted to the particle-soliton hybrid and continues to slide on the surface of the hybrid structure after they are in direct contact (Figs. 6b, c), indicating that the moving toron is not yet in its free energy minimal state. In the final equilibrium configuration, this toron is observed to be bond to the handlebody particle and has an elliptical cross-section, which reveals modifications of the original axially symmetric director arrangement of this particle-like solitonic object. In contrast to the isotropic repulsions between torons reported previously [60], the symmetry-breaking interactions with the handlebody particle induce anisotropic morphing of the toron's director profile in the homeotropic cell's confinement. While our observations here reveal just one example of such complex interaction, they point into the interesting opportunities for exploring self-assembly of topologically complex handlebodies with different orientations relative to the far-field director and symmetry axes of solitonic objects like torons and skyrmions.

**3.3 Nonreciprocal dynamics of director and motions of soliton-particle hybrid structures**

Hybrid structures in CNLCs exhibit intriguing dynamic behaviors under an AC electric field (Figs. 7, 8). These motions, including squirming-like translational and rotational dynamics, arise due to symmetry-breaking effects in the surrounding LC medium's director field. With the system being in the low Reynolds number regime, the inertia effects associated with colloidal particles in LCs are negligible. Therefore, the observed dynamics arise from the complex balance of viscous, elastic, and (when the field is applied) dielectric torques. The asymmetry of the medium's response during the field-on and field-off periods of different voltage driving schemes is a key contributor to the nonreciprocal responses of the oscillating fields [31] This nonreciprocal director field evolution can then result in motions of both solitonic and colloidal objects [63] individually or in large assemblies like schools. The present study probes yet another case of motion of hybridized soliton-particle assemblies.

Applying an AC electric field to hybrid structures generates frequency-dependent nonreciprocal dynamics of the LC director field, as evidenced by the motions of $g=3$ and $g=1$ handlebody particles (Figs. 7, 8). For example, translational motion is observed for a $g$=3 handlebody particle with homeotropic anchoring, dispersed into a homeotropic cell, when an AC field with peak-to-peak voltage $U_{pp}$=8 V at the frequency of 3 and 4 Hz is applied (Fig. 7a). The mean squared displacement extracted from tracking the spatiotemporal trajectory of the particle increases with time (Fig. 7f), yet the deviation from the linear dependency suggests that the motion is more than purely diffusive.[64] The dramatic difference in the particle motions compared between frequencies of the voltage driving changing from 3 to 4 Hz shows how sensitive the system is to even modest changes in the voltage driving scheme. As the frequency exceeds 5 Hz, the translational motion turns into rotational motion (Fig. 7b), with the angular speed increasing as frequency increases (Fig. 7g). For the $g$=1 handlebody, rotational motion is observed as well (Fig. 7c), but the angular speed decreases as frequency increases in this case (Fig 7h).

While a standalone toron structure is axially symmetric, hybridization between a toron and a $g$=1 handlebody can break such symmetry and lead to electricity-driven rotational motion. At a relatively higher frequency, 1 kHz where the solitonic structure remains stable, the toron-handlebody hybrid rotates with a periodicity of approximately 10 s, which decreases as the voltage increases between $U_{pp}$=5-6 V (Figs. 7d, e, and i). This soliton-colloidal system's behavior is different from the one depicted in Fig. 7c, despite both involving a $g$=1 handlebody particle. In Figs. 7d, e, and i, the particle does not rotate about its symmetry axis but revolves around the central part of the toron. In fact, the presence of the toron and its complex hybridization with the particle-induced deformations is crucial for producing such rotational behavior. Indeed, if the AC electric field is increased with the applied peak-to-peak voltage over $U_{pp}$=6.2 V, leading to the solitonic structure disappearance (Fig. 8), the rotational motion of the handlebody particle ceases. Despite the stable and smooth rotation of the hybridized structure at lower voltages, the standalone particle stays stationary at $U_{pp}$=6.2 V and higher (Fig 8c), with this observation supporting the notion that soliton-mediated symmetry breaking plays an essential role in driving the studied emergent dynamics.

## 4 Discussion

The chirality of CNLCs gives origins to the stability of various particle-like topological solitons. Colloidal spheres, previously studied in CNLC media, have been found to hybridize with such topological solitons, like hopfions and torons [65-68]. Enriched with the presence of colloidal inclusions with nontrivial surface topology, the system we study in the present work provides a unique testbed for exploring the interplay between the topologies of director fields and surfaces of rigid colloidal objects. Our study reveals how the interplay between these different types of topologically nontrivial objects in chiral host media gives origins to a wide variety of fascinating emergent phenomena, ranging from different types of hybridization to bound states. The behavior is in many ways more complex than what was observed in the past for colloidal spheres hybridizing and self-assembling with topological solitons in CNLCs under similar geometric confinement [66-68]. Colloidal surfaces induce singular topological defects that obey mathematical theorems describing relations between topological invariants characterizing surfaces and defects, similar to what was previously found for other nematic host media [14-16,67]. For example, topologically trivial surfaces such as $g=0$ spheres in a nematic LC typically induce two $s=1$ surface boojums for tangential boundary conditions, thus satisfying the constraint $\Sigma s=2$. For $g=1$ handlebodies, however, the constraint becomes $\Sigma s=0$, which does not require extra defects on the surface [16], though self-compensating defects can appear to minimize free energy. This indeed was the case when such particles with planar surface anchoring were placed in a nematic LC, where it does not induce any singular defects or induces self-compensating oppositely-charged pairs of surface defects that have topological invariants adding to the net zero winding number [16, 67]. While the chiral nature of the CNLCs, in addition, enriches the physical behavior of colloidal inclusions via hybridization with solitonic features, such general conclusions related to singular defects remain unchanged both for colloidal handlebodies studied here and, as was found elsewhere, for colloidal spheres [66, 67]. To comply with the topological constraints and minimize the free energy costs, the spheres are often found replacing the hedgehog defects of a toron or being fully enclosed by the spatially-localized triple-twist director structure when small enough. For handlebody particles, the hybridization includes incorporating them into solitonic structures (Fig. 1), or wrapping the particles with solitonic loops (Fig. 2). The topological constraints are still satisfied in both cases, despite the more complex director configurations of hybrid structures. In the first case, the $g=1$ handlebody with planar anchoring does not require nor induce any defects, while the triple-twist toron structure is compensated by the ½ disclination loop and the hedgehog defect. In the second case, the hopfion structure is topologically stable without needing extra defects, and the handlebody induces two disclination loops with opposite signs such that the net topological charge stays equal to 0.

While this study focuses on hybridization in commonly used nematic host media, such as 5CB and CB7CB, the use of low birefringence LCs in future studies will allow for improved FCPM imaging, enabling a more precise reconstruction of director configurations in hybridized solitonic structure.

The effect of chiral symmetry-breaking in the medium with orientational elasticity and localized solitonic structures also manifests itself in the attractive interaction between the handlebody particles as well as their nonreciprocal motions under AC electric fields, including squirming-like translational and rotational dynamics. These motions arise from the interplay between the anisotropic properties of LCs and interactions with solitons, which induce director field evolution in response to the electric field's on-off symmetry breaking in the system. Intriguingly, even at high frequencies where the particles can no longer fully respond to changes in the electric field, nonreciprocal rotational motion persists. While the exact mechanism behind this high-frequency emergent behavior remains unclear, the pivotal role of solitons in enabling it has been confirmed by our experiments, as the rotation is not observed when solitons are not present, highlighting their critical role in this phenomenon. Further investigations are anticipated to elucidate the underlying principles driving this dynamic response and to explore its potential applications in advanced micro-robotics and active colloidal systems.

**Conclusions**

This study has revealed the complex but beautiful interplay between topological solitons, singular defects, and colloidal handlebody particles co-hosted in CNLCs. We have demonstrated how torons, cholesteric finger loops, and other solitonic structures interact with topologically complex colloidal particles, leading to hybridized particle-soliton configurations, elasticity-mediated interactions, and ensuing bound states, as well as nonreciprocal dynamics yielding translational and rotational motions. The hybridized states could be further reconfigured by electric fields, allowing control over the spatial arrangement of solitonic features and enabling transitions between different director configurations and colloidal dynamics. Periodic oscillations in the electric field have induced nonreciprocal dynamics, such as squirming-like translational and rotational motions of the handlebody particles. Such nonreciprocal behaviors not only open new possibilities for controlling motions and assembly in colloidal systems but also suggest pathways for precise particle manipulation and energy-efficient active transport at the micrometer-to-millimeter scales.

Overall, our study paves a foundation for future research on pre-designed chiral metamaterials that can be reconfigured using topological and elastic interactions in LCs. Soliton-colloid interactions and nonreciprocal motions due to symmetry-breaking effects promise to deepen the interdisciplinary links between topology and colloidal soft matter systems while also potentially enabling applications in soft chiral metamaterials, microrobotics, and optoelectronic devices. Our studies presented here are non-exhaustive in terms of topologies of solitons and colloidal particles, as well as possible hybridized states that they can form and dynamics that they can exhibit, which we extend to expand and generalize in future studies. Additionally, it will be of interest to extend our work to soft matter media with polar orientational order, like LC helielectrics [69] and chiral ferromagnetic colloids [70], as well as to the regime of nanoscale colloidal inclusions [71,72].

**Acknowledgements:** We are grateful to our former team members: Q. Liu for fabricating handlebody particles [14-16] and P. Ackerman for the code [31,32] used in the director field reconstruction. We acknowledge technical assistance and discussions with B. Senyuk. Y.Y. acknowledges the financial support from the Japan Society for the Promotion of Science (JSPS KAKENHI grant number JP24K23088). I.I.S. acknowledges hospitality of Hiroshima University during his sabbatical stay, when this work was initiated, as well as the financial support of the fundamental research on mesostructured colloidal self-assembly by the U.S. Department of Energy, Office of Basic Energy Sciences, Division of Materials Sciences and Engineering, under Award ER46921, contract DE-SC0019293 with the University of Colorado in Boulder.

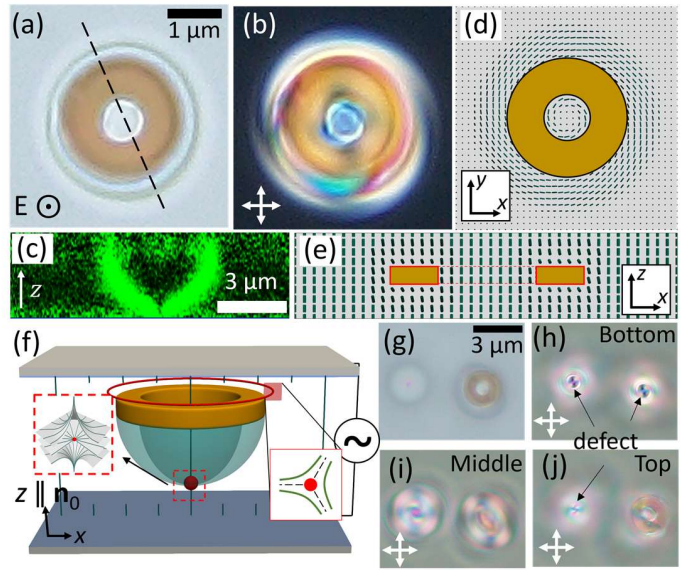

**Fig. 1** Hybrid structure of a *g*=1 handlebody particle, bound with a toron in a chiral nematic medium under an electric field. (a) Bright-field microscopy and (b) polarized optical microscopy (POM) images showing the particle embedded in the solitonic region, with a distinct boundary indicating the hybrid structure. The circled dot symbol marked with E indicates the direction of the applied electric field. (c) Cross-sectional FCPM image along the black dashed line in (a) with polarization direction along the *x*-direction. The particle has planar anchoring and is immersed into a CNLC with a pitch of 10 μm; the particle-LC suspension is constrained within a 5-μm-thick cell with homeotropic boundary condition. (d) In-plane and (e) cross-sectional schematic diagrams of the director configuration around a *g*=1 handlebody particle with planar anchoring embedded in a homeotropic far-field director. Dark green cylinders represent the LC director, while the brown torus shows the *g*=1 handlebody particle's position and orientation. (f) Schematic diagram of the hybrid structure with the red dot indicating the hedgehog defect and the red line marking the line defect. Insets to the left and right depict the director distribution around the point defect and a cross-sectional schematic of the disclination line forming a closed loop, respectively. (g) Bright-field microscopic image of the hybridized configuration (right) and a standalone toron (left) next to each other. (h-j) POM images captured at different focal planes within a cell with a chiral pitch and cell thickness of 20 μm. The crossed white arrows represent the polarization directions of the polarizer and the analyzer.

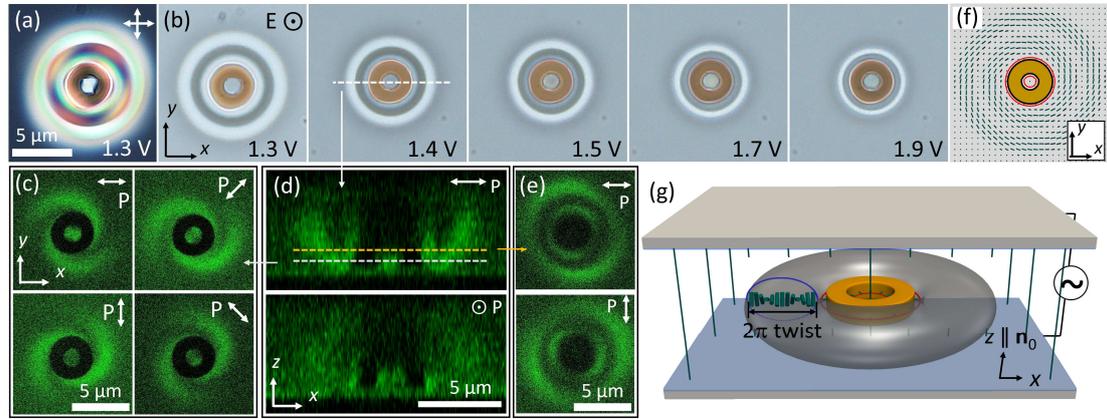

**Fig. 2** Hybridized structure of a $g$=1 handlebody particle and a topological soliton under an electric field. (a) POM image of a hybridized configuration of a $g$=1 handlebody and a cholesteric finger-loop-like solitonic configuration. The particle, suspended in a CNLC with a chiral pitch of 10 μm, has the homeotropic anchoring conditions and is confined within a 5-μm-thickness homeotropic cell. (b) Bright-field images under varying electric fields. The circled dot symbol marked with E indicates the direction of the applied electric field. (c) FCPM cross-sectional images of the *xy*-plane with polarization angles of 0°, 45°, 90°, and 135° relative to the *x*-axis, obtained along the white dotted line shown in panel (d). (d) Cross-sectional FCPM images for excitation light's linear polarizations parallel and perpendicular to the cross-section plane, respectively. (e) FCPM images along the *xy*-plane with polarization directions of 0° and 90° relative to the *x*-axis, obtained along the yellow dotted line in panel (d). (f) Schematic diagram of the director configuration around the hybrid structure in the *xy*-plane, where cylinders indicate the LC director orientations and the brown torus represents the handlebody particle. (g) Schematic diagram of the hybrid structure, with a gray solitonic structure wrapping the handlebody particle, illustrating the 2π twist in the radial direction. The red lines inside and outside of the handlebody particle represent half-integer disclination loops equivalent to that shown in Fig. 1d.

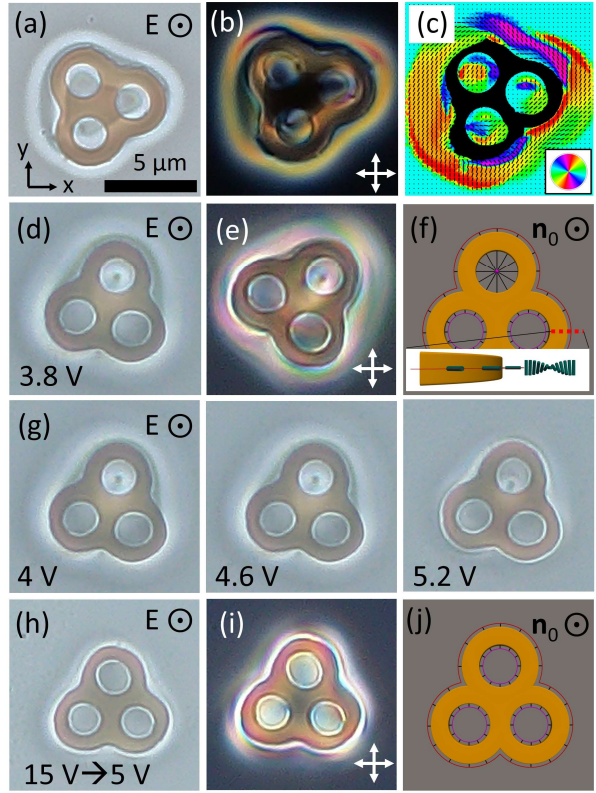

**Fig. 3** Hybrid structures of a *g*=3 handlebody particle and solitonic configurations. (a) Bright-field and (b) POM images showing the *g*=3 particle with planar anchoring, embedded in a solitonic structure. (c) The reconstructed director field of the complex hybrid structure in the *xy*-plane containing the particle. (d) Bright-field and (e) POM images of a *g*=3 handlebody with homeotropic anchoring, showing a point defect within the solitonic region. (f) Schematic diagram of the *g*=3 handlebody with a point defect; the inset illustrates twist director deformation from the particle's outer surface to the bulk. (g) Bright-field microscopy images with varying electric field magnitude, showing that the point defect moves toward the center of the particle, and the solitonic structure shrinks as voltage increases. (h) Bright-field and (i) POM images of the hybrid structure of a *g*=3 handlebody with the homeotropic anchoring boundary conditions on its surface after increasing voltage above $U_{pp}$=15 V and decreasing back to $U_{pp}$=5 V. The point defect in (g) opens into a disclination loop. (j) Schematic diagram of the *g*=3 handlebody containing only line defects closed into loops. Red lines and magenta lines in (f) and (j) indicate disclination lines forming loops of opposite $\pi_2(\mathbb{S}^2/\mathbb{Z}_2)$ hedgehog charge, respectively. The magenta sphere in (f) indicates a point defect of *m*=+1.

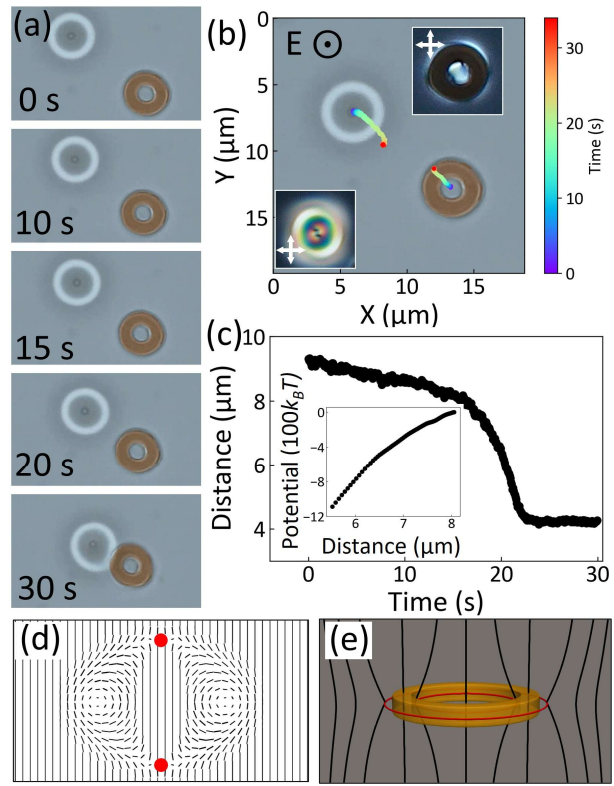

**Fig. 4** Attractive elasticity-mediated interaction between a toron and a g=1 handlebody particle. (a) Bright-field images showing the gradual approach of a toron and a handlebody particle over time. (b) Color-coded trajectories of the toron and particle versus time, with insets displaying POM images of each structure. The direction of an applied electric field is marked on the image. (c) Plot of the distance between the toron and particle as a function of time. The inset shows the relation between potential energy and distance calculated from the data in the distance versus time plot. (d) Schematic diagram of the director configuration in the vertical cross-section of a toron. (e) Schematic diagram of the director configuration around the handlebody particle with the disclination loops both in the exterior and inside the hole. The black and red lines indicate the director field and disclination line, respectively.

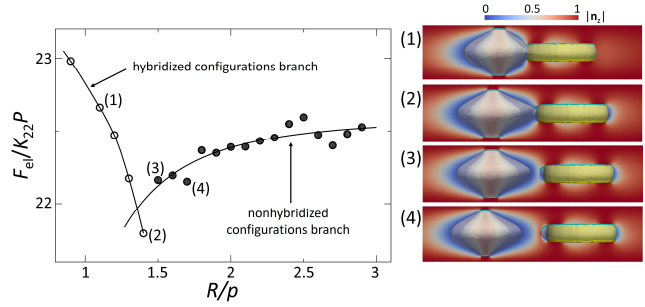

**Fig. 5** Elastic part of the Landau-de Gennes free energy versus separation distance between a toron and a $g=1$ handlebody with homeotropic anchoring. The circles represent the results of the numerical minimization of the Landau-de Gennes free energy, and the solid curves are a guide to the eye. LC configurations at the right correspond to the toron-handlebody separations marked by (1)-(4) on the free energy curves. The LC configurations are visualized using the color-coded $z$-component $|\mathbf{n}_z|$ of $\mathbf{n}(\mathbf{r})$. The transparent grey surfaces correspond to the isosurfaces $|\mathbf{n}_z|=0.05$, and the blue tubes around handlebodies and at the top and bottom of the torons enclose the regions of the reduced scalar order parameter $Q \leq 0.7 Q_b$ and correspond to the disclination line defects, where $Q_b$ is the bulk value of the scalar order parameter.

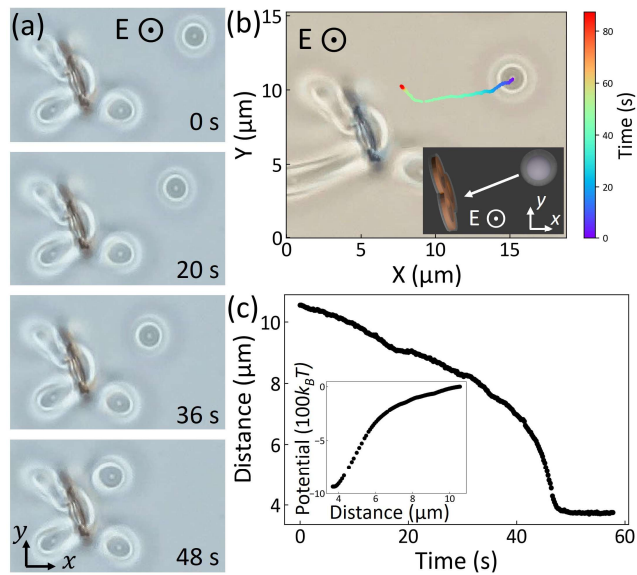

**Fig. 6** Attractive interaction between a toron and a hybrid structure consisting of a $g=3$ handlebody particle decorated by solitonic structures. (a) Bright-field images showing the toron approaching the hybrid structure. The $g=3$ handlebody with planar anchoring is aligned perpendicularly to the confining substrates. (b) Color-coded trajectory of the toron. The inset is a schematic diagram of this behavior. (c) Plot of the distance between the toron and the particle as a function of time; the inset shows the dependency of potential energy on distance.

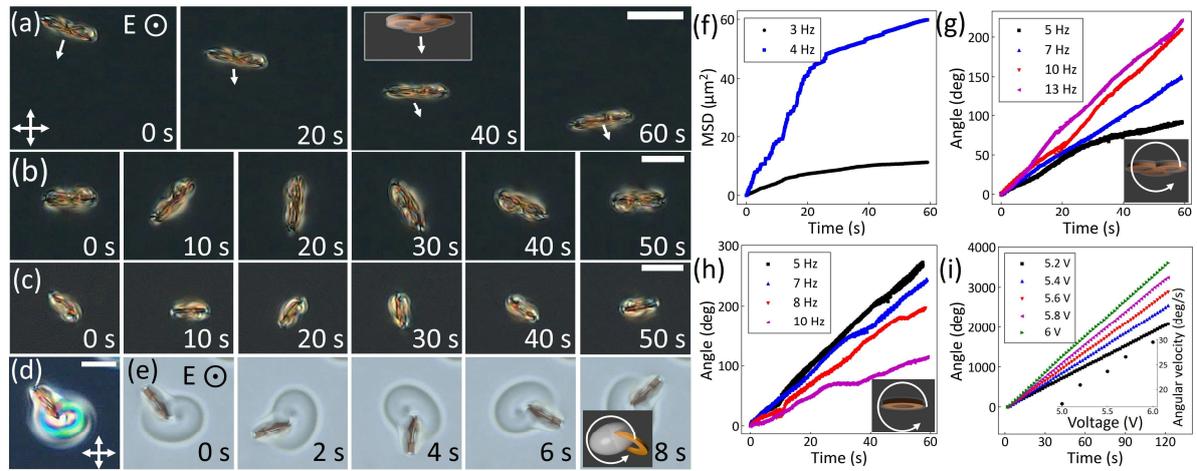

**Fig. 7** Dynamics of *g*=3 and *g*=1 handlebody particles undergoing rotational and translational motions in CNLCs. (a) POM images capturing the translational motion of a *g*=3 handlebody over time under the square-wave AC electric field of 8 V and 3 Hz. (b) POM images of rotational motion of the same *g*=3 handlebody for the AC electric field of 13 Hz. (c) POM images of rotational motion of a *g*=1 handlebody particle over time for the electric field of 8 V and 5 Hz. (d) POM image of a hybrid structure of a *g*=1 handlebody particle and a toron. The scale bars in (a)-(d) each represent 5 μm. (e) Bright-field microscopy images showing the rotational dynamics of the hybrid structure in (d), observed over time for an electric field of frequency 1 kHz and amplitude of 5.2 V. (f) Plot of the mean squared displacement of the translational motion of the *g*=3 handlebody in (a). (g, h) Plots of the rotational angle versus time for a *g*=1 handlebody in (b), *g*=3 handlebody in (c), and hybrid structure in (d), respectively. The inset in (i) shows the plot of angular speed versus the applied voltage. The insets in (a), (e), (g), and (h) show schematic diagrams of the moving objects, i.e., handlebody particles and hybrid structures. The white arrows indicate the direction of the translational motion in (a) and the more complex rotational-translational dynamics in others. The second mixture mentioned in the method section was used for experiments of images (a-c) and corresponding plots (f-h).